\documentstyle[12pt,equation]{article}
\begin{document}
\def\be{\begin{equation}}
\def\ee{\end{equation}}
\newcommand{\een}{\end{subequations}}
\newcommand{\ben}{\begin{subequations}}
\def\bea{\begin{eqnarray}}
\def\eea{\end{eqnarray}}
\def\phad{\mbox{$P_{\rm had}$}}
\def\ptmin{\mbox{$ p_{T,min} $}}
\def\gamgam{\mbox{$\gamma \gamma $}}
\def\xg{\mbox{$x_{\gamma}$}}
\def\totwo{$2 \rightarrow 2$}
\def\glph{\mbox{${ f_{g/\gamma}(x,Q^2)}$}}
\def\qvph{\mbox{${ \vec q^{\gamma} } $}}
\def\qph{\mbox{${q^{\gamma} } $}}
\def\vqxqsq{\mbox{$ \vec q^{\gamma} (x,Q^2)  $}}
\def\vqisq{\mbox{${ f_{q_{i}/\gamma} (x,Q^2) }$}}
\def\vqisqm{\mbox{$ q_i^{\gamma} (x,Q^2)$ }}
\def\gamg{${ \gamma g }$}
\def\gamgam{\mbox{$\gamma \gamma $}}
\def\gamp{\mbox{$\gamma p$}}
\def\sigtot{\mbox{$\sigma^{tot}_{\gamp}$}}
\def\jetgamp{\mbos{$\sigma^{jet}_{\\gamp}$}}
\def\eplem{\mbox{$e^+ e^- $}}
\def\rts{\mbox{$ \sqrt{s} $}}
\def\qg{\mbox{$q^{\gamma}$}}
\def\fGg{\mbox{$f_{G/{\gamma}$}}}
\def\xqsq{\mbox{$(x,Q^2)$}}
\def\fge{\mbox{$ f_{\gamma/e}$}}
\def\qsq{\mbox{$Q^2$}}
\def\nav{\mbox{$\langle n_{\rm jet} \rangle $}}
\def\f2gam{\mbox{$ F_2^\gamma $}}
\renewcommand{\thefootnote}{\fnsymbol{footnote}}

\pagestyle{empty}
\begin{flushright}
BU TH-94/01 \\
hep-ph/9407205 \\
\end{flushright}
\vspace*{1cm}
\begin{center}
{\Large \bf Structure of  photon and the muon puzzle.}
\footnote{Talk
presented at the International Conference on Non-accelerator Particle
Physics, Bangalore, January 2-9, 1994.}
\\
\vspace*{5mm}
Rohini M. Godbole \\
\vspace*{5mm}
{\it Physics Dept., Bombay Univ., Vidyanagari, Santa Cruz (East),
Bombay 400098, India} \\
\end{center}
\begin{abstract}
In this talk I first briefly explain the puzzle presented by the high
muon content observed in air-showers which point to the ultra
high-energy $\gamma$ sources like the HER-X-1. Since it had been
suggested that a possible explanation of the puzzle might come from
the effects of the hadronic structure of the photon, I briefly explain
the concept of photon structure function and comment on the
uncertainties in the predictions of the \sigtot. Then I show that
while the current experiments at the \eplem\ (TRISTAN and LEP) and
$ep$ (HERA) colliders have seen clear evidence for the hadronic
structure of the photon, the observed muon excess in the air shower
experiments, if confirmed by other experiments to be at the same high
level, can not be explained in terms of the photon structure function.
\end{abstract}

\section{What is $\mu$ puzzle?}
One of the aims of the TeV/PeV $\gamma$ ray astronomy is to gain
information about the origin of cosmic rays \cite{prep}. The extensive
air--shower array experiments look for point sources of ultrahigh
energy $\gamma$ rays \cite{kiel,tonwar,dingus,SPASE}.  These
experiments can serve as directional cosmic $\gamma$ ray
telescopes. But they have to be able to distinguish between photon and
hadron initiated air--showers. One of the criteria that is normally
used is the expected low $\mu$ content of the $\gamma$ showers.  The
$\gamma$ initiated showers are supposed to be $\mu$ poor and the
reasoning goes as follows. The muons in the air--showers come from
$\pi \rightarrow \mu \nu $ as well as from the Bethe-Heitler
production of $\mu$ pairs and at still higher energies from the heavy
quark decays.  In case of $\gamma $ induced air--showers, the total
cross--section is dominated by the Bethe-Heitler production of $e^+e^-$
pairs.  Hence one expects the  $\mu$ content of the $\gamma$
initiated showers to be at  few $\%$ level that of (roughly a factor 30
below) the hadron initiated showers.

However, so far there never has been evidence for directional showers
associated with point sources of $\gamma$ rays which are $\mu$
poor. On the other hand, the experiments \cite{dingus}
reported air showers associated with known `point' $\gamma$
sources whose $\mu$ content is consistent with that of the
hadron initiated showers or even more. Ref. \cite{kiel} reported  observation
of air showers associated with the point source Cyg X-3,  ref.
\cite{tonwar} reported air--showers associated with Crab-Nebula.
 Air showers reported in ref. \cite{dingus} were associated with the point
source HER X-1 which were $\mu$ rich, {\it i.e.}, their $\mu$ content
was consistent with that of a hadronic shower.
This observation as well as any failure to see a $\mu$ poor shower
associated with `point' sources,  signalled existence of what is termed as
the $\mu $ puzzle, {\it i.e.}, the cosmic ray air shower experiments
`see' more muons than they `ought' to.

Let us now turn more specifically to  the theoretical
predictions \cite{gaisser,manhik,burrows,halzen5} of the $\mu$ content of
the $\gamma$ induced showers. The actual predictions involve
detailed Monte Carlo simulation of the shower developement, but
the essential features of how  one arrives at these predictions
can be summarised as follows (see,  e.g., ref. \cite{manhik}). One
starts with measured photoproduction cross--sections in the
laboratory. In the days before the results from the $ep$
collider HERA at DESY, the maximum centre of mass energy
$(\sqrt{s})_{\gamp}$ for which data were available was $\sim 20$
GeV. At these energies the  \sigtot\ is $\sim \; 100  \; \mu b
$. Now to estimate the $\mu $ content of the airshowers one has
to extrapolate these cross--sections to TeV/PeV energies using
the experimentally well known logarithmic rise of total
cross--sections. Using this one then calculates the production
of hadrons and hence production of muons, telling us how many
$\mu'$s ought to be there in the $\gamma$ induced showers. The
observed $\mu$ excess \cite{dingus} would require a
\sigtot\ $\sim \; O(100)\; mb$ for PeV energy photons. This
would imply  a faster than the assumed logarithmic growth of
\sigtot\ in going from the  laboratory  measurements at GeV
energies to the PeV energies involved in the cosmic ray
experiments. This can happen only if there exist some new
threshold in the photonuclear ($\gamma $ air ) cross--sections
which gives it a steeper energy dependence. Hadronic structure
of the high energy $\gamma$ (in particular the gluon
content) can indeed cause a sharp increase in the
cross--sections with increasing $\gamma$ energy \cite{manhal}.
It should be noted here that  the increasing importance of the
hadronic structure of the $\gamma$ with rising energy was not
{\it introduced} here to `explain' the $\mu$ `puzzle', but is a
prediction of perturbative QCD. What is not clear, and hence has
been a topic of debate, is how much does the hadronic structure of
photon   contribute  to the rise in \sigtot\ with energy.

\section{Hadronic content of photon}
The terminology of the `structure ' of a photon is essentially a
short hand way of describing how a high energy photon interacts
with other particles: hadrons and photons.  The idea that
photons behave like hadrons when interacting with other hadrons
dates back to the early days of strong interaction physics and
is known to us under the name of the Vector Meson Dominance
(VMD) picture. This essentially means that at low 4--momentum
transfer, the interaction of a photon with hadrons is dominated
by the exchange of vector mesons which have the same quantum
numbers as the photon.  While this picture works reasonably well
for `soft' processes (i.e., reactions characterized by small
4--momentum transfer), it is not at all clear that it should
describe the whole story of interactions of photons with hadrons
at high energies as well. In the VMD picture one then expects
$$
\sigma_{\gamma p} ^{tot} \propto \alpha \; \sigma_{Vp}^{tot} ,
$$
where $\alpha $ is the  fine structure constant.
However, since the photon `behaves' like a hadron while
interacting with other hadrons it must be possible to get
information about the photon structure just like the other
hadrons, e.g., the proton. This information is obtained by
studying the deep inelastic scattering (DIS) of high energy
leptons of energy E off proton targets,
\be
e^- + p \rightarrow e^- + X
\label{dis}
\ee
The double differential  cross--section  for the process is a
function of two independent variables $y = \nu /E $  where
$\nu $ is the energy carried by the probing photon in the laboratory
frame, and $x=Q^2 /( 2 M \nu ) $ where M is the proton
mass and $ - Q^2 $ is the invariant mass of the virtual photon
\begin{figure}[hbt]
\vspace{4cm}
\caption{Deep Inelastic Scattering for the proton and photon.}
\label{disfig}
\end{figure}
in fig. \ref{disfig}a. In the quark-parton-model (QPM) it is
given by,
\be
\frac{d^2\sigma^{ep \rightarrow X}}  {dx dy} = \frac {2 \pi\
\alpha^2 \ s}
{Q^4} \times \bigg[(1+(1-y)^2)\ F_2^p (x) - y^2\ F_L^p(x) \bigg],
\label{discs}
\ee
where
\bea
 F_2^p(x) &  = &  \sum_{q} e_q^2 \  x\ f_{q/p}(x); \nonumber \\
 F_L^p(x) & = & F_2^p (x) - 2 x F_1^p(x) \nonumber
\eea
are the two electromagnetic structure functions of the proton
(in the QPM $F_L^p(x)$ is identically zero but not so in QCD) and
$ f_{q/p}(x) $  the probability for quark $q$ to carry a momentum
fraction $x$ of the proton and $e_q$ denotes the electromagnetic
charge of quark $q$ in units of the proton charge.

To measure the structure function of a photon such an
experimental situation is provided at \eplem\ colliders in
$\gamma^* \gamma $ reactions as shown in fig.~\ref{disfig} (b).
Here the virtual photon with invariant mass square $-Q^2 $
probes the structure of the real photon. If the VMD picture were
the whole story then one would expect that such an experiment
will find
\be
F_2^\gamma \simeq F_2^{\gamma,VMD} \propto F_2^{\rho^0} \simeq F_2^{\pi^0}.
\label{f2VMD}
\ee
Then with increasing \qsq,
the  structure function \f2gam\ will behave just like a hadronic
proton structure function and shrink to lower values of $x$ as
predicted by QCD \cite{glap}. However, there is a very important
difference in case of photons,{\it i.e.}, photons possess pointlike
couplings to quarks. This has interesting implications for $\gamma^* \gamma $
interactions as first noted in the framework of the
QPM by Walsh \cite{wal}.  It
essentially means that $\gamma^* \gamma $ scattering in
fig.~\ref{disfig} contains two contributions as shown in
\begin{figure}[hbt]
\vspace{4.5cm}
\caption{Two contributions to \f2gam.}
\label{fgamtwo}
\end{figure}
fig.~\ref{fgamtwo}. The contribution of fig.~\ref{fgamtwo}~(a) can be
estimated by eq.(\ref{f2VMD}), whereas that of fig.~\ref{fgamtwo}~(b)
was calculated in the QPM~\cite{wal}. This is done by considering
the cross--section for the reaction
\begin{displaymath}
\gamma + \gamma^* \rightarrow q + \bar q.
\end{displaymath}
Due to $t$ and $u$ channel poles this can be calculated only when one considers
quarks with finite masses. The result can  be recast in a form equivalent to
eq. (\ref{discs}):
\bea
{ {d^2\sigma^{e\gamma \rightarrow X}} \over {dx dy}} &= &{ {2
\pi
\alpha^2  s_{e \gamma}} \over {Q^4}} \times {3 \alpha \over \pi}
\nonumber \\
& & {  \sum_q e_q^4 \bigg\{ (1+(1-y)^2)\times {\bf [}x(x^2
+(1-x)^2) \times \ln{{W^2} \over {m_q^2}}} \nonumber \\
& & +{ 8x^2 (1-x) -x{\bf ]} - y^2 {\bf [} 4 x^2 (1-x)
{\bf ]} \bigg\}},
\label{qpm}
\eea
where $ W^2 = Q^2 (1-x)/ x.$
On comparing  eqs.(\ref{discs}) and (\ref{qpm}) we see that the
factors in square brackets in the above equation  have the
natural interpretation as photon structure functions $ F_2^\gamma $
and $F_L^\gamma $ and one has
\bea
F_2^{\gamma,\rm{pointlike}}(x,Q^2) &= & 3 {\alpha \over \pi}
                            { \sum_q e_q^4 \bigg[ x (x^2 +(1-x)^2)
                             \times
\ln{{W^2} \over {m_q^2}} + 8x^2(1-x) -x \bigg]} \nonumber \\
                       & = & {\sum_q e_q^2\;  x\
                       f_{q/\gamma}^{\rm{pointlike}}(x,Q^2)}.
\label{qpmpred}
\eea
Two points are worth noting: the function $
F_2^{\gamma,\rm{pointlike}}(x,Q^2) $ can be completely calculated in
QED and secondly this contribution to \f2gam\ increases
logarithmically with \qsq. So in this simple `VMD + QPM' picture,
\f2gam\ consists of two parts, $F_2^{\gamma,\rm{pointlike}}$ and $
F_2^{\gamma,\rm{VMD}}$, with distinctly different \qsq\ behaviour and
with the distinction that for one part both the $x$ and the \qsq\
dependence can be calculated completely from first principles.

This QPM prediction received further support when it was shown by
Witten \cite{wit77} that at large \qsq\ and at large $x$, both the $x$
and \qsq\ dependence of the quark and gluon densities in the photon
can be predicted completely even after QCD radiation is included.  An
alternative way of understanding this result is to consider the
evolution equations \cite{dewit} for the quark and gluon densities
inside the photon. These contain an inhomogeneous term on the r.h.s
proportional to $\alpha$, which describes $\gamma \rightarrow q
\bar{q}$ splitting, i.e. the pointlike coupling of photons to
quarks. In the `asymptotic' limit of large \qsq\ and large $x$, the
\vqisq\ have the form
\bea
f_{q_i/{\gamma}}^{\rm asymp}(x,Q^2)& \propto & {\alpha \times
\ln \left( {Q^2}
\over {\Lambda^2_{\rm{QCD}}} \right) F_i(x) }\nonumber \\
&\simeq & {{\alpha \over {\alpha_s}} F_i(x)},
\label{asymp}
\eea
where $\rm \Lambda_{QCD} $ is the usual QCD scale parameter,
$\alpha_s (Q^2) $  is given in terms of the running strong coupling
constant by $g_s^2 (Q^2) \over 4 \pi $ and the $x$
dependence of the $F_i(x)$ is completely calculable. Note here
the factor  $\ln \left( {Q^2} \over {\Lambda^2_{\rm{QCD}}}
\right)$ on the r.h.s.  Measurements \cite{exrev} of the photon
structure function $F_2^{\gamma}$ in $\gamma^* \gamma$ processes
did indeed confirm the basic QCD predictions of the linear rise
of $F_2^\gamma$ with $\ln \left(Q^2\right)$ at large $x$.
This discussion thus means that just like one can `pull' quarks
and gluons out of a proton one can look upon the photon as a
source of partons and that the parton content of the photon
rises with its energy. Physically this means that the photon
splits in a $ q \bar q$ pair and these radiate further gluons
and thus fill up a volume around photon with partons.

The asymptotic solutions discussed above, though very useful to
understand the rise of the photon structure function with \qsq\ , are
valid only at large $x$ and large \qsq . At small values of $x$ these
solutions diverge, indicating thereby that `hadronic' part of \f2gam\
can not be neglected at small $x$. Hence in practice, this separation
of $F_2^\gamma$ in two parts as in fig.~\ref{fgamtwo} is not very
meaningful, especially when one wants to use this parton language to
predict the high energy photon interactions.  Although the debate on
the subject is not yet closed \cite{comment} , it is now generally
accepted that it is better to forego the absolute predictions of
\f2gam\ of the asymptotic part, that are possible in perturbative QCD
(pQCD) and use only the prediction of the \qsq\ evolution of the
photon structure function in analogy to the case of the proton
structure function. At present there exist eight different
parametrisations of the photon structure function
\cite{comment,comment1}. The DIS measurements described above measure
only the quark-parton densities \vqisq\ (for $x > 0.05 $ and $Q^2 <
100-200$ GeV$^2$) directly and \glph\ is only inferred indirectly. As
a result there is considerable uncertainty in the knowledge of
\glph. The different parametrisations differ quite a lot from each
other in the gluon content. It should also be mentioned here, that
these differences reflect the differences in different physical
assumptions in getting \glph\ from the data on \f2gam. So independent
information on \glph\ is welcome.

\section{ Calculation of jet production in \gamgam,
\gamp\ collisions}
In this section let us discuss how one can compute the high
energy $\gamma$ cross--sections using the parton language and
what are the crucial factors affecting these predictions. In the
parton language, interaction  of a hadron with others can be
described, at high energies and  for processes involving final
state particles at large angles to the original beam direction
(large transverse momentum $p_T$), in terms of the scattering of
the  pointlike constituents  inside the  two hadrons against
each  other.  In this picture, the differntial cross--section
for the production of a pair   of two large $p_T$ jets in the
collision of a $\gamma$ with a proton(say) will be given by
\bea
\label{generic}
{d \sigma \over dp_T}   (\gamma p \rightarrow {\rm jets} \;  +
{\rm X} )& = & \sum_{P_1,P_2,P_3,P_4}
\int dx_{\gamma}\; f_{P_{1}/\gamma} (x_{\gamma})
\int dx_p \; f_{P_{2}/p}(x_p) \nonumber   \\
& &\quad\qquad\times {d \hat \sigma \over dp_T} \left(P_1 + P_2 \rightarrow
P_3  + P_4\right),
\eea
where the  sum is over all the different intital (final) state
partons $P_1, P_2$ ($P_3,P_4$)  and $d \hat \sigma \over dp_T $ is the
subprocess cross--section that can be computed in pQCD. As we
have already seen in the discussions on \f2gam\   one can get
effectively `real' high energy $\gamma$    beams in the
laboratory by using $e$--beams and then making sure that the
final state  $e$ is scattered in the forward direction at a
very small angle. In this situation the large $p_T$ jet
production in high energy $e-p$ collisions  can be computed as
\bea
\label{ep}
{d \sigma \over dp_T}   (e p \rightarrow {\rm jets} \;  +
{\rm X} )& = & \sum_{P_1,P_2,P_3,P_4} \int dz\; f_{\gamma/e}(z)
\int dx_{\gamma} \; f_{P_{1}/\gamma}(x_{\gamma}) \nonumber \\
& &\qquad\quad\times\int dx_p\; f_{P_{2}/p}(x_p)\;\;
{d \hat \sigma \over dp_T} \left(P_1 + P_2 \rightarrow  P_3  + P_4\right).
\eea
The $\gamma$  induced processes are of two types :
\begin{enumerate}
\item[{1)}] The `direct'
processes, an example of which is depicted in
\begin{figure}[hbt]
\vspace{4cm}
\caption{`Direct' and `resolved' contributions to jet production in
$ep$ collision.}
\label{fig}
\end{figure}
fig.~\ref{fig}~(a),
where the $\gamma$ couples directly to the partons in the
photon. In this case all the energy of the   photons  goes into
the subprocess and and the hadronic content of the $\gamma$
plays no role and  in  eq.(\ref{ep}),
$$
f_{P_{1}/\gamma}(x_{\gamma}) = \delta \; (1 - x_{\gamma}).
$$
\item[{2)}]The `resolved' processes  where the partons in the proton
interact with  the partons in the photon and hence only a
partial fraction of the $\gamma $ energy is available for the
subprocess. One of the possible contribution is shown in fig.
\ref{fig} (b).
\end{enumerate}
It should be noted here that
although the `resolved' processes have an extra factor of $\alpha_s$,
due to the factor of ${\displaystyle {\alpha \over {\alpha_s}}}$
in the parton desities  of the photon (recall eq.
(\ref{asymp})), both the `direct' and the `resolved' processes are
formally of the same order in the coupling constants.
With rising $\gamma$ energies, increasingly  more energy
becomes available for the subprocess, at a fixed $p_T$ or mass
of the final state. Hence  the importance of the `resolved' processes
increases with the $\gamma$ energies.  The gluon content
of  photon \glph\ is peaked at small values of $x$   and
hence the importance of the contributions involving the gluon
in the photon  in  the initial state in  eq. (\ref{ep}), increases
with the increasing $\gamma$ energy at a fixed $p_T$ or with
decreasing $p_T$ at a fixed $\gamma$ energy. The  `resolved'
events will also have  additional `spectator' jets in the direction
of the incident $\gamma$ ({\it i.e.} the direction of the
incident  $e$).

Such high photon energies are available at the HERA collider at
DESY (Hamburg) in the collision of a 30 GeV $e$ beam with a $p$
beam of 800 GeV. This corresponds to c.m. energies for the \gamp\
system $\leq 300 $ GeV (which in turn means $E_\gamma \leq 40 $
TeV in the frame where the proton is at rest). Indeed a
calculation \cite{dgprd} showed that at the HERA collider   the
photo--production of jets in the process
$$
\gamma + p \rightarrow {\rm jets} + X ,
$$
is dominated by the `resolved' processes upto $p_T \simeq 40 $
\begin{figure}[hbt]
\vspace{7cm}
\caption{Ratio of resolved and direct contributions for
$d \sigma (ep \rightarrow$ jets$)/  d p_T$ as a function of $p_T$
\protect\cite{dgprd}. For details see ref. \protect\cite{dgprd}.}
\label{junk}
\end{figure}
GeV. This dominance is sensitive to the gluon content of the
photon and fortunately not very sensitive to the choice of gluon
parametrisation of the proton,  partially because of the  much
better knowledge of $f_{g/p}$  in the relevant $x_p$ region. An
example  is shown in fig. \ref{junk} for two different
parametrisations of \glph\ that were then availble. Since then
the HERA experiments H1 and ZEUS \cite{H1,ZEUS} have studied the
photo-production of the jets, and  confirmed  the existence of
the `resolved' contribution at the expected level,  verified
various expected qualitative features of the resolved
contributions such as existence of the spectator jets in the
backward direction, different angular distribution expected for
the jets produced from the hard scattering of the partons in the
photon etc. An example of the same is shown in
\begin{figure}[hbt]
\vspace{6cm}
\caption{The total $ep$ cross--section measured \protect\cite{ZEUS}
for transverse energies larger than $E_T^0$. The curve is the HERWIG
prediction, using the DG parametrization with \ptmin\ = 1.5 GeV.}
\label{figfive}
\end{figure}
fig.~\ref{figfive}. As a matter of fact the `resolved' contributions to the
jet--production  have been isolated and  used to extract \glph\
\cite{talk}.

Similar studies of the jet--production in \gamgam\ collisions
\cite{amylep} at the \eplem\ collider TRISTAN and LEP, have
confirmed the existence of the `resolved' contributions
\cite{tristanac}.  These studies have already ruled out some of
the very hard parametrisations  of $\glph$ \cite{amylep,uehara}.
Thus these observations have  provided a confirmation
(in addition to the DIS measurements ) of the ideas about \f2gam\
and these experiments will continue to add to our knowledge of
the \glph , \vqisq .

\section{\f2gam\ and QCD prediction of \sigtot}
What is more relevant for the issue of $\mu$ puzzle is the
total photoproduction  cross--section \sigtot . The calculation
of the `hard' processes such as the jet--production
cross--sections is well defined in pQCD but   valid  upto
$\simeq $ 1-2 GeV.  The total inclusive \gamp\ cross--section
for production of jets with $p_t > p_T^{\rm min}$ given by,

\be
\label{incl}
\sigma^{\gamma p}_{\rm incl}  = \int_{p_{T,min}} {d\sigma
\over dp_T} (\gamma p \rightarrow  j_1  +  j_2 + X\;)\; dp_T,
\ee
rises very strongly with decreasing $p_{T,min}$ at a fixed
$\gamma$ energy or with $\gamma$ energy at a fixed $p_{T,min}$.
The differential cross--section ${d\sigma
\over dp_T} (\gamma p \rightarrow  j_1  +  j_2 + X\;)$ receives
contribution from the `direct' as well as the `resolved'
processes as discussed before and it also rises strongly with
decreasing $p_T$.
At high \gamp\ c.m. energies the resolved photon processes then
cause copious production of the `minjets'  and  a very rapid
rise of the inclusive \gamp\  cross--section, as was first pointed
out in ref. \cite{manhal}.  An example is shown in
\begin{figure}[hbt]
\vspace{6cm}
\caption{Predictions \protect\cite{manhal} of the increase of the
inclusive (mini)jet cross--section in $\gamma p$ collisions with
\protect\rts, for \protect\ptmin\ = 2 GeV and
various parametrizations for $f_{\vec q/{\gamma}}$.}
\label{incjet}
\end{figure}
fig.~\ref{incjet}, for \ptmin\ = 2 GeV and various parametrizations of
$f_{\vec q/{\gamma}}$.
Of course, the total cross--section cannot grow indefinitely at the rate
shown in fig.~\ref{incjet}; some mechanism will have to unitarize it.
This problem is well known for hadronic ($pp$ or $p \bar{p}$) collisions.
In this case unitarization is usually achieved by eikonalization. The
crucial observation here is that LO QCD predictions for cross--sections,
like those shown in fig.~\ref{incjet}, refer to {\em inclusive} jet
cross--sections; in other words, they differ from the jet production
contribution to the total cross--section by a factor of the average jet
pair multiplicity \nav.
Formally one writes (following ref. \cite{nuceik} and modifying
for the $\gamp$ case as pointed out in \cite{collad})
\be \label{gp}
\sigma_{\gamma p}^{\rm inel} = \int d^2 b\; \phad \left\{ 1 - exp \left[ -
\left( \sigma_{\gamma p}^{\rm hard}(s) + \chi_{\gamma p}^{\rm soft} \right)
A(b)/\phad \right] \right\},
\ee

Here $\vec{b}$ is the two--component impact parameter, $A(b)$
describes the transverse overlap of partons in the nucleon and the
photon, $\sigma_{\gamma p}^{\rm hard}$ is the perturbative QCD
prediction for the minijet cross--section (obtained by integrating $d
\sigma / d p_T$ in the region $p_T \geq \ptmin$), and $\chi_{pp}^{\rm
soft}$ is the non--perturbative (soft) contribution to the eikonal,
which is fitted from low--energy data and \phad\ is a parameter
describing the probability that the photon goes into a hadronic state;
clearly $\phad \sim {\cal O}(\alpha)$.  Thus we see that unlike the
predictions of the jet cross--sections the predictions of \sigtot\
depend not only on the \glph\ and \vqisq\ but also on $p_{T,min}$,
\phad , $A(b)$ and the nonperturbative contribution to the eikonal.
There is considerable theoretical uncertainty in the ans\"atze uses
for $\sigma^{\rm soft}$ and the choice of \phad , as well as
$p_{T,min}$ and $A(b)$ \cite{storrow,ina,flet4}.  Thus the predictions
of the total \gamp\ cross--sections will depend on all these
parameters apart from the information about
\glph\ and \vqisq.

Thus to recapitulate the  predictions for \sigtot\  in the
frameowrk of QCD depend on the following factors:
\begin{enumerate}
\item The parton densities in the photon, particularly \glph .
The  range of $x$ values that are relevant shifts to smaller values
from $x < 0.01 $ to $x < 0.001$ or smaller, as one goes from
HERA energies to the case of the PeV energy photons relevant for
the $\mu$ puzzle.
\item Value of $p_{T,min}$.
\item Modelling of the soft cross--section.
\item The probability \phad\ for the proton  to go to a hadron state.
\item The transverse overlap $A(b)$ of the partons in the nucleon and
the photon.
\end{enumerate}

Fortunately the experiments at HERA \cite{zeustot,h1tot} also measured
\sigtot\ in addition to electro-(photo-)production of `hard'
jets. This information from HERA when coupled with information about
the multi-jet production in \gamgam\ collisions \cite{amylep,uehara},
can be used to reduce the uncertainty in the knowledge of \glph\ and
$p_{T,min}$.  The \gamgam\ data are consistent with $p_{T,min} \simeq
2.0 - 2.5 $ GeV depending upon the choice of parametrisation for
\glph\ and \vqisq.  But the very broad gluon distribution of LAC3
parametrisation is ruled out already by these data. The observation of
the resolved contribution to large $p_T$ jets at HERA
\cite{H1,ZEUS,talk} allows an extraction of \glph\ which is
essentially consistent with almost all the modern parametrisations
(that rise somewhat steeply at small $x$)  other than LAC3
and a similar value of $p_{T,min}$. It should also be mentioned here
that the present `extractions' of \glph\ has large experimental as
well as (Monte Carlo related) theoretical errors. Both these errors
are likely to go down in further analysis/data from HERA, allowing
perhaps a better discrimination.

As seen above, calculation of \sigtot\ involves, in addition to
\ptmin\ and \glph , \vqisq,  \phad ,  $A(b)$ as well as
model for $\sigma^{\rm soft}$.  Here some model--builders take
recourse to the low--energy measurements of \sigtot\ to determine the
model parameters. Here almost the whole cross--section is dominated by
the soft process and the only relevant `hard' contribution is the
`direct ' process.  As a result of these the low energy predictions
are independent of $f_{\vec q /{\gamma}}$ and can determine some of
the nonperturbative parameters, especially $\chi^{\rm soft}$ quite
well. The ZEUS measurement \cite{zeustot} along with the (pre-HERA)
predictions using the eikonalised calculation mentioned above for
different choices of the parameters mentioned are shown in
\begin{figure}[hbt]
\vspace{7cm}
\caption{Comparison of various predictions of total $\gamma p$ cross-sections
with low--energy data and the recent ZEUS measurement \protect\cite{zeustot}.}
\label{gptot}
\end{figure}
fig.~\ref{gptot}.  The two solid curves show fits to low--energy data
based on Pomeron phenomenology which does not involve any `hard'
contribution.  The two dot--dashed curves show minijet
predictions \cite{schuler} using the DG parametrization with
\ptmin\ = 1.4 (upper) and 2.0 (lower curve) GeV, while the
dotted and dashed curves have been obtained from the LAC1
parametrization using the same values of \ptmin. The LAC
parametrization seems disfavoured but in view of the theoretical
and experimental uncertainties it might be premature to exclude
it altogether. The DG minijet prediction with \ptmin\ = 2 GeV is
certainly in agreement with the data.  At the HERA energy the
effect of eikonalisation on the predictions for \sigtot\ is not
very large, but it is certainly substantial when the PeV energy
$\gamma$ are involved. A further clarification between different
models (eikonalised minijet  or Pomeron based) will be possible
if one can measure energy--dependence of \sigtot. At present,
one can say definitely:
\begin{enumerate}
\item The resolved contribution to  photoproduction of jets does
rise with increasing $\gamma$ energies and has been seen.
\item These processes also  cause the \sigtot\  to rise with energy.
However, with \ptmin\ $\simeq\ 2$ GeV and almost all the current
parametrisations of \glph\ and \vqisq\ this rise is much less steep
than the most `optimistic' pre-HERA predictions of \sigtot.
\item  Further measurements of jet cross--sections, details of
event shapes, e.g,  multiplicity and various correlations \cite{nugamgen}
should provide further help in clarifying this situation.
\end{enumerate}

Armed with this information now one can turn towards the
predictions of \sigtot\ for the TeV/PeV energy photons from the
point $\gamma$ sources.
\begin{figure}[hbt]
\vspace{8.5cm}
\caption{$\sigma_{\rm inel}^{\gamma \; {\rm air}}$ and
$\sigma_{\rm inel}^{\gamma \; p}$ cross--section as a
function of $\gamma$ energy in an eikonalised  model.
\protect\cite{ina}.}
\label{showercsec}
\end{figure}
Fig.~\ref{showercsec} shows predictions of a particular model
\cite{ina} for $\sigma_{\rm inel}^{\gamma \;{\rm air}}$. This model of
\gamp\ interactions fits the \ptmin\ from $\pi \; N $ scattering data
and the quark-gluon densities of the `hadronic states' into which the
photon converts itself are related to parton-distributions for a
pion. The model predictions, with the inclusion of soft scattering
background, agree with data at low-energy (8 GeV $\leq
\sqrt{s}_{\gamma p} \leq 20 $ GeV ) quite well and are reasonable
(though somewhat high) for HERA energies.  As can be seen from this
figure $\sigma_{\rm inel} $ does rise with increasing $E_\gamma$ (or
equivalently $\sqrt {s}_{\gamp}$) and is about $\simeq 2-3 $ mb for
the PeV energy photons. This falls way short of the $\sigma \simeq $
100 mb required to explain the muon excess observed in the CYGNUS
experiment. One should keep in mind that since \nav\ does rise sharply
with energy (recall fig. \ref{incjet}) it is possible that
fluctuations could sometimes produce $\mu$ rich showers. But if the
future air shower experiments continue to see showers associated with
point sources, with $\mu$ content as high as that seen by the earlier
experiments \cite{dingus}, the explanation must lie somewhere else and
not with the rising \gamp\ cross--sections

So in conclusion we can say that
\begin{enumerate}
\item There exists a  hint of a $\mu$ excess seen  by  the
extensive air shower experiments in the air showers associated
with `point' sources. This would mean  that the $\gamma$
induced showers  look more  like hadronic showers.
\item Since the photon does behave like a hadron at high energies
and the `resolved' contribution of the `partons in the photon'
increases with $\gamma$ energy, feature (1) could have been, in
principle,  due to the hadronic content of $\gamma$.
\item The current collider experiments at  \eplem\ (effective
\gamgam)  colliders  TRISTAN and LEP and $ep$ (effective \gamp)
collider HERA have seen these contribiutions  coming from
the   hadronic structure of the photon and have provided some
information about \glph.
\item The prediction of  \sigtot\  has  considerable theoretical
uncertainties, but using the  current experimental  information from
the \eplem, $ep$ colliders on  jet  production and on \sigtot\
from  HERA, one can conclude that the hadronic structure of
$\gamma$  can cause an increase in \sigtot\ by a factor of  $\simeq$
2-3 for the PeV energy photons but not much more.
\item Hence the   photon strucure function effects can give rise
to $\mu$ rich showers only as a fluctuation but if future
experiments continue to  report    existence of extremely $\mu$
rich showers ,  the   explanation  can not be    provided    in
terms of the photon structure.
\end{enumerate}
\subsection*{Acknowledgements}
I wish to thank M. Drees for an enjoyable  and  long standing
 collaboration  during which some of the   results presented
in this talk were obtained.


\begin{thebibliography}{99}
\bibitem{prep} T.C. Weekes, Phys. Reports, {\bf C 160},1 (1988)
and references therein;\\
For a review see also, e.g., R. Protheroe, in {\it Proceedings
of the 20th International Cosmic Ray Conference}  Moscow, 1987, eds
V.A. Kozyzrivsky et al., (Nauka, Moscow, 1987) {\bf vol. 8}.
\bibitem{kiel}  M. Samorski and W. Stamm, Ap. J. ,
{\bf 268}, L17 (1983).
\bibitem{tonwar} Acharya et al., Nature {\bf 347}, 3614 (1990).
\bibitem{dingus} B.L. Dingus et al., Phys. Rev. Lett. {\bf 61},
1906 (1988).
\bibitem{SPASE} J. van Stekelenborg  et al., Phys. Rev. {\bf D 48},
4495 (1993).
\bibitem{gaisser} T. Stanev, T.K. Gaisser and F. Halzen, Phys.
Rev. {\bf D 32}, 1244 (1985).
\bibitem{manhik}  M. Drees, F. Halzen and K. Hikasa, Phys. Rev.
{\bf D 39}, 1310 (1989).
\bibitem{burrows} R. Gandhi, I. Sarcevic, A. Burrows, L. Durand
and H. Pi, Phys. Rev. {\bf 42}, 290 (1990).
\bibitem{halzen5} T.K. Gaisser, T. Stanev, F. Halzen, W.L. Long and E.
Zas, Phys. Rev. {\bf D 43}, 314 (1991).
\bibitem{manhal} M. Drees  and F. Halzen, Phys. Rev. Lett. {\bf
61}, 275 (1988).
\bibitem{glap}
G. Altarelli and G. Parisi, Nucl. Phys. {\bf B126}, 298 (1977);\\
V.N. Gribov and V.N. Lipatov, Sov. J. Nucl. Phys. {\bf B 15}, 78 (1972).
\bibitem{wal}
T.F.~Walsh, Phys. Lett. {\bf B36}, 121 (1971) ;\\
S.M.~Berman,J.D.~Bjorken and J.B.~Kogut, Phy. Rev. {\bf D4}, 3388 (1971) ;\\
P.~Zerwas and T.~Walsh, Phys. Lett. {\bf B44}, 195 (1973) .
\bibitem{wit77}
E.~Witten, Nucl. Phys. {\bf B120}, 189 (1977) .
\bibitem{dewit}
R.J.~deWitt, L.M.~Jones, J.D.~Sullivan, D.E.~Willen
and H.W.~Wyld Jr.,  Phys. Rev. {\bf D19}, 2046 (1979).
\bibitem{exrev}
For an early review, see,
Ch. Berger and W. Wagner, Phys. Rep. {\bf 146}, 1 (1987);\\
J. Olsson, Nucl. Phys. B, Proc. Suppl. {\bf 3}, 613 (1988);\\
For a collection of more recent experimental information see, e.g.,
Proceedings of the {\it IX International Workshop on Photon--Photon
collisions, San Diego, California, March 1992.}
\bibitem{comment}
For a more complete discussion of the subject see, {\it e.g.},
M. Drees and R.M. Godbole, Pramana {\bf 41}, 83 (1993) and references
therein.
\bibitem{comment1}
Since this talk was given, the number of these parametrisations has
now increased from 8 to 14 due to newer parametrisations given by
K. Hagiwara, T. Izubuchi, M. Tanaka and I. Watanabe in  KEK-TH-{\bf
376}.
\bibitem{dgprd}
M. Drees and R.M. Godbole, Phys. Rev. {\bf D 39}, 169 (1989).
\bibitem{H1}
H1 collaboration, T. Ahmed et al., Phys. Lett. {\bf B 297}, 205 (1992);
\bibitem{ZEUS}
ZEUS collaboration, M. Derrick et al., {\bf B 297}, 404 (1992).
\bibitem{talk}
H1 collaboration, T. Ahmed et al., Phys. Lett. {\bf B 314}, 436
(1993);\\
ZEUS collaboration, M. Derrick et al., {\bf B 322}, 287 (1994).
\bibitem{amylep}
AMY collaboration, Phys. Lett. {\bf B 277}, 215 (1992);
Phys. Lett. {\bf B 325}, 248 (1994);\\
TOPAZ collaboration Phys. Lett. {\bf 314}, 149 (1993);\\
ALEPH collaboration, Phys. Lett. {\bf B 313}, 509 (1993).
\bibitem{tristanac}
M. Drees and R.M. Godbole, Nucl. Phys. {\bf B339}, 355 (1990).
\bibitem{uehara}
For an experimental review of the subject see, e.g., S. Uehara, KEK
Preprint 93-112.
\bibitem{nuceik}
T.K. Gaisser and F. Halzen, Phys. Rev. Lett. {\bf 54}, 1754 (1985);\\
L. Durand and H. Pi, Phys. Rev. Lett. {\bf 58}, 303 (1987).
\bibitem{collad}
J.C. Collins and G.A. Ladinsky, Phys. Rev. {\bf D43}, 2847 (1991).
\bibitem{storrow}
J.R. Foreshaw and J.K. Storrow, Phys. Lett. {\bf B 321}, 159 (1994).
\bibitem{ina}
K. Honjo, L. Durand, R. Gandhi, H. Pi and I. Sarcevic,  Phys.
Rev. {\bf D 47}, R4815 (1993); Phys. Rev. {\bf D 48}, 1048 (1993).
\bibitem{flet4}
 R.S. Fletcher, T.K. Gaisser and F. Halzen, Phys. Lett. {\bf 298}, 442
(1993).
\bibitem{zeustot}
ZEUS Collab., M. Derrick et al., Phys. Lett. {\bf B 293} (1992) 465;
DESY 94-032.
\bibitem{h1tot}
H1 Collab., T. Ahmed et al., Phys. Lett. {\bf B 299}, 374 (1993).
\bibitem{schuler}
G. Schuler, Proc. of the Workshop on HERA physics, DESY 1991,
editors W. Buchm\"uller and G. Ingelmann.
\bibitem{nugamgen}
G. Schuler and T. Sjostrand, Nucl. Phys. {\bf B 407}, 539 (1993);
Phys. Rev.  {\bf D 49}, 2257 (1994).
\end{thebibliography}
\end{document}